\documentclass[prl,twocolumn,aps,amsmath,nofootinbib,superscriptaddress]{revtex4}

\usepackage{graphicx}
\usepackage{bm}
\usepackage{epsfig}

  \newcount\hour \newcount\minute
  \hour=\time \divide \hour by 60
  \minute=\time
  \newcommand{\mydate}{\ \today \ - \number\hour :\ifnum \minute<10 0\fi 
\number\minute}
\def\OMIT#1{}

\newcommand{\nn}{\nonumber}

\newcommand{\bea}{\begin{eqnarray}}
\newcommand{\eea}{\end{eqnarray}}

\begin{document}

%%%%%%%%%%%%%%%%%%%%%%%%%%%%%%%%%%%%%%%%%%
%Define Title, Author, Address, Preprint#

\preprint{ \hbox{MIT-CTP 3620} \hbox{CMU-HEP-0504} \hbox{UCSD-PTH-05-04}
 \hbox{hep-ph/xxxxxxx}  }

\title{\boldmath 
A Precision Model Independent Determination of $|V_{ub}|$ from $B\rightarrow \pi l \nu$ \\
%\ \mydate 
}
\author{M.~Christian Arnesen\vspace{0.3cm}}
\affiliation{Center for Theoretical Physics, Massachusetts Institute of
  Technology, Cambridge, MA 02139}
\author{Ben Grinstein}
\affiliation{Department of Physics, University of California, San Diego,
La Jolla, CA, 92093}
\author{Ira Z.~Rothstein}
\affiliation{Department of Physics, Carnegie Mellon University,
    Pittsburgh, PA 15213  \vspace{0.3cm}}
\author{Iain W. Stewart\vspace{0.1cm}}
\affiliation{Center for Theoretical Physics, Massachusetts Institute of
  Technology, Cambridge, MA 02139}
%%%%%%%%%%%%%%%%%%%%%%%%%%%%%%%%%%%%%%%%%%
\begin{abstract}
   
  A precision method for determining $|V_{ub}|$ using the full range in $q^2$ of
  $B\to \pi \ell\nu$ data is presented.  At large $q^2$ the form
  factor is taken from unquenched lattice QCD, at $q^2=0$ we impose a model
  independent constraint obtained from $B\to \pi\pi$ using the soft-collinear
  effective theory, and the shape is constrained using QCD dispersion relations.
  We find $|V_{ub}| =(3.54\pm 0.17\pm  0.44) \times 10^{-3}$. With 5\%
  experimental error and 12\% theory error, this is competitive with inclusive
  methods.  Theory error is dominated by the input points, with negligible
  uncertainty from the dispersion relations.

\end{abstract}

\maketitle 

The remarkable success of the B-factories have lead to a new era for precision
results in the CKM sector of the standard model. For $|V_{ub}|$ inclusive and
exclusive measurements from semileptonic decays should yield a precise value,
but must surmount the now dominant theoretical uncertainties.  For inclusive
decays measuring $|V_{ub}|$ is more difficult than $|V_{cb}|$ because cuts make
observables either sensitive to a structure function which demands input from
radiative decays, or require neutrino reconstruction.  The heavy flavor
averaging group (HFAG)'s average from inclusive decays based on operator product
expansion techniques is $10^{3}|V_{ub}| =4.7 \pm 0.4$~\cite{HFAG}.  Exclusive
techniques for $|V_{cb}|$ use heavy quark symmetry (HQS) to normalize the form
factors.  For $|V_{ub}|$ from $B\to\pi\ell\bar\nu$ symmetry techniques fall
flat, and model independent form factor information relies on precision lattice
QCD.

Recently, the Fermilab~\cite{FNAL} and HPQCD~\cite{HPQCD} groups have presented
unquenched lattice results for $B\to\pi$ form factors.  Uncertainties in the
discretization restrict the kinematics to pions that are not too energetic
$E_\pi\lesssim 1\,{\rm GeV}$, which for the invariant mass of the lepton pair is
$15\,{\rm GeV}^2 \lesssim q^2 \le 26.4\,{\rm GeV}^2$.  Unfortunately, since the
phase space goes as $|\vec p_\pi|^3$, there are less events and more
experimental uncertainty in this region. For $\bar B^0\to \pi^+\ell\bar\nu$
\begin{eqnarray} \label{rate}
  {d\Gamma}/{dq^2} &=& 
     N\: |V_{ub}|^2\: |\vec p_\pi|^3\ 
    |f_+(q^2)|^2 \,,
\end{eqnarray}
where $N=G_F^2/(24\pi^3)$.  For example, Belle~\cite{Belle} found
\begin{eqnarray} 
 10^3\,|V_{ub}|_{q^2\ge 16} =\! \bigg\{\! \begin{array}{l}
 3.87\pm 0.70\pm 0.22^{+0.85}_{-0.51} \mbox{ (FNAL)} \\[4pt]
  4.73\pm 0.85\pm 0.27{}^{+0.74}_{-0.50} \mbox{ (HPQCD)} 
 \end{array} 
\end{eqnarray}
where the errors are statistical, systematic, and theoretical. In quadrature
this is an uncertainty of $\sim 25\%$. 

The latest Babar, CLEO, and Belle average is~\cite{Dingfelder},
\begin{eqnarray} \label{avg}
  \mbox{Br}(\bar B^0\to\pi^+\ell^-\bar\nu) = (1.39\pm 0.12)\times 10^{-4} \,,
\end{eqnarray}
which should yield $|V_{ub}|$ at the $\simeq 5\%$ level. So far extractions of
$|V_{ub}|$ from the total Br rely on QCD sum rules~\cite{Ball} and quark models
for input.  For example, HFAG reports results on ${\rm Br}(B\to
\{\pi,\rho,\omega\}\ell\bar\nu)$ that lead to central values $10^{3}|V_{ub}|
=2.9$ to $3.9$~\cite{HFAG}. Due to the uncertainty they do not currently average
over exclusive extractions of $|V_{ub}|$.

In this letter we present a model independent exclusive method for determining
the entire $B\to \pi$ form factor $f_+(q^2)$ and thus $|V_{ub}|$. A total
uncertainty $\delta |V_{ub}|\simeq 13\%$ is achieved by combining 1) the
unquenched lattice results~\cite{FNAL,HPQCD}, 2) a constraint at $q^2=0$ derived
from SCET~\cite{Bpipi} and $B\to\pi\pi$ data, which determines $|V_{ub}|
f_+(0)$, and 3) dispersion relations and analyticity which allow us to
interpolate over the entire region of $q^2$ by bounding the shape of $f_+(q^2)$
between input points~\cite{Bourrely,Lebed}. The SCET constraint induces an
additional implicit functional dependence on $|V_{ub}|$ in the form factors. Our
first analysis uses just the total Br, yielding an analytic formula for
$|V_{ub}|$.  The second includes $q^2$-spectra with a $\chi^2$ minimization
which allows the experimental data to constrain the theoretical uncertainty. A
different approach for including the $q^2$-spectra was developed
in~\cite{Onogi} based on the Lellouch distribution method~\cite{Lellouch}.

{\em Analyticity Bounds.}
We briefly review how analyticity constrains the $B\to \pi$ form factors, $f_+$
and $f_0$, referring to~\cite{Bourrely,Lebed,Boyd} for more
detail. Our notation follows~\cite{Boyd}, and we set $t_\pm=(m_B \pm m_\pi)^2$.
Suitable moments of a time ordered product of currents, $\Pi^{\mu\nu}(q^2) = i
\int\!\!  d^4x \, e^{iqx} \langle 0 | T J^\mu(x) J^{\dagger\nu}(0)|0\rangle$ can
be computed with an OPE in QCD and are related by a dispersion relation to a
positive definite sum over exclusive states
\begin{eqnarray} \label{Pimunu}
  {\rm Im}\, \Pi^{\mu\nu}\!\! =\!\! \int\!\![\mbox{p.s.}]\, 
  \delta(q\!-\!p_{B\pi})
   \langle 0 |
  J^{\dagger\nu}|\bar B \pi \rangle \langle \bar B \pi | J^\mu |0\rangle 
  + \ldots 
\end{eqnarray}
Keeping this first term bounds a weighted integral over $t_+\le t\le\infty$ of
the squared $B\pi$ production form factor. Using analyticity and crossing
symmetry this constrains the shape in $t=q^2$ of the form factors for $B\to \pi$
in the physical region $0\le t\le t_-$. The results are simple to express by
writing each of $f_+(t)$, $f_0(t)$ as a series
\begin{equation}
\label{ff}
   f(t)=\frac{1}{P(t) \phi(t,t_0)}\sum_{k=0}^\infty a_k(t_0)\:  z(t,t_0)^k \,,
\end{equation}
with coefficients $a_k$ that parameterize different allowed functional forms.
The variable
\begin{equation}
   z(t,t_0)=\frac{\sqrt{t_+-t}-\sqrt{t_+-t_0}}{\sqrt{t_+-t}+\sqrt{t_+-t_0}},
\end{equation}
maps $t_+< t<\infty$ onto $|z|=1$ and $-\infty < t< t_+$ onto $z\in [-1,1]$.
$t_0$ is a free parameter that can be chosen to attain the tightest possible
bounds, and it defines $z(t_0,t_0)=0$. We take $t_0 = 0.65\, t_-$ giving $-0.34
\le z\le 0.22$ for the $B\to \pi$ range. In Eq.~(\ref{ff}) the ``Blaschke''
factor $P(t)$ eliminates sub-threshold poles, so $P(t)=1$ for $f_0$, while
$P(t)=z(t;m_{B^*}^2)$ for $f_+$ due to the $B^*$ pole.  Finally, the
``outer'' function is given by
\begin{eqnarray}
 \phi(t,t_0)\!\!\! &=&\!\!\! 
 \sqrt{\frac{n_I}{K \chi_J^{(0)}}}
 \big(\sqrt{t_+\!-\!t}\!+\!\sqrt{t_+\!-\!t_0}\big)
  \frac{(t_+ \!-\! t)^{(a+1)/4}}{(t_+ \!-\! t_0)^{1/4}}
  \nn \\
  && \hspace{-1.5cm}\times \!
  \big(\sqrt{t_+\!-\!t}\!+\!\sqrt{t_+}\big)^{-(b+3)}
 \big(\sqrt{t_+\!-\!t}\!+\!\sqrt{t_+\!-\!t_-}\big)^{a/2} 
 \,,
\end{eqnarray}
where $n_I=3/2$ and for $f_+$: $(K=48\pi,a=3,b=2)$, while for $f_0$:
$(K=16\pi/(t_+ t_-),a=1,b=1)$. Here $\chi_J^{(0)}$ corresponds to the lowest
moment of $\Pi(q^2)$ computed with an OPE.  At two loops in terms of the pole
mass and condensates and taking $\mu=m_b$~\cite{Gir,Lellouch}
\begin{eqnarray} \label{chi}
 \chi_{f_+}^{(0)}
 \!\!&=&\!\!  \frac{3\big[ 1\! +\! 1.140\, \alpha_s(m_b)\big]}{32\pi^2m_b^2} 
  \!-\! \frac{\overline{m}_b\:\langle  \bar u u\rangle}{m_b^6}
  \!-\! \frac{\langle \alpha_s G^2\rangle}{12\pi m_b^6} \,,
  \nn\\
 \chi_{f_0}^{(0)}
 \!\! &=&\!\! \frac{\big[ 1 \!+\! 0.751\,  \alpha_s(m_b)  \big]}{8\pi^2} 
 + \frac{\overline{m}_b\: \langle  \bar u u\rangle}{m_b^4}
  + \frac{\langle \alpha_s G^2\rangle}{12\pi m_b^4}
 \,,\ \ \ \ 
\end{eqnarray}
with $\overline{m}_b \langle \bar u u\rangle \simeq -0.076\,{\rm
  GeV}^4$, $\langle \alpha_s G^2 \rangle \simeq 0.063 {\rm GeV}^4$. We use
$m_b^{\rm pole}=4.88\,{\rm GeV}$ as a central value. With
Eq.~(\ref{ff}) the dispersive bound gives a constraint on the coefficients
\begin{equation}
  \label{sum}
  \sum_{k=0}^{n_A} a_k^2 \leq 1 \,,
\end{equation}
for any choice of $n_A$.

Eqs.~(\ref{ff}) and (\ref{sum}) give only a weak constraint on the normalization
of the form factor $f_+$. In particular, data favors $a_0\sim 0.02$, so
$a_0^2\ll 1$. The main power of analyticity is that if we fix $f_+(q^2)$ at
$n_A$ input points then it constrains the $q^2$ shape between these points.
With $n_A=5$ the error from the bounds is negligibly small relative to other
uncertainties, as we see below (our analysis is also insensitive to the exact
values of $\chi_J^{(0)}$ or $m_b$). The bounds can be strengthened using heavy
quark symmetry or higher moments of $\Pi(q^2)$~\cite{Boyd}, but since this
uncertainty is very small we do not use these improvements.

{\em Input Points.}
A constraint at $q^2=0$ is useful in pinning down the form factor in the small
$q^2$ region.  Here we implement a constraint at $q^2=0$ on $|V_{ub}| f_+(0)$
that follows from a $B\to\pi\pi$ factorization theorem derived with
SCET~\cite{Bpipi}.  The result holds in QCD and uses isospin symmetry and data
to eliminate effects due to the relative magnitude and strong phase of penguin
contributions.  Manipulating formulas in~\cite{Bpipi} we can write the result in
terms of observables
\begin{eqnarray} \label{Vf}
   |V_{ub}| f_+(0) \!\!&=&\!\! \bigg[ \frac{64\pi }{ m_{B}^3 f_\pi^2 }\frac{\overline Br(B^-\to\pi^0\pi^-)}{\tau_{B^-} |V_{ud}|^2 G_F^2} \bigg]^{1/2}\\
 && \hspace{-1.3cm}
 \times \bigg[ \frac{(C_1+C_2)t_c -C_2}{C_1^2-C_2^2}\bigg] \bigg[1 + {\cal
   O}\Big(\alpha_s(m_b), \frac{\Lambda_{\rm QCD}}{m_b}\Big)\bigg]\,,\nn
\end{eqnarray}
where $C_1=1.08$ and $C_2=-0.177$ are parameters in the electroweak Hamiltonian
at $\mu=m_b$ (we drop the tiny $C_{3,4}$), and $t_c$ is a hadronic parameter
whose deviation from $1$ measures the size of color suppressed amplitudes. In
terms of the angles $\beta,\gamma$ of the unitarity triangle and CP-asymmetries
$S_{\pi^+\pi^-}$ and $C_{\pi^+\pi^-}$ in $B\to\pi^+\pi^-$,
\begin{eqnarray} \label{tc}
 t_c \!\!&=&\!\!
  \sqrt{\overline R_c\: \frac{(1\!+\!   B_{\pi^+\pi^-}\cos2\beta
   +S_{\pi^+\pi^-} \sin2\beta)}{2 \sin^2\!\gamma}} \,,
\end{eqnarray}
with $\overline R_c\!=\! [{\overline Br(B^0\to\pi^+\pi^-)\tau_{B^-}}]/[2
{\overline Br(B^-\to\pi^0\pi^-)\tau_{B^0}}]$, and $B_{\pi^+\pi^-} \!=\! (1\!-
C_{\pi^+\pi^-}^2\!-\!  S_{\pi^+\pi^-}^2)^{1/2}$.  Eqs.~(\ref{Vf},\ref{tc})
improve on relations between $B\to\pi\pi$ and $B\to\pi\ell\bar\nu$ derived
earlier, such as in Ref.~\cite{BBNS}, because they do not rely on expanding in
$\alpha_s(\sqrt{m_b\Lambda})$ or require the use of QCD sum rules for input
parameters to calculate $t_c$.

Using the latest $B\to\pi\pi$ data~\cite{HFAG}, Eq.~(\ref{Vf}) gives
\begin{equation} \label{f0}
 \qquad\quad f^0_{\rm in} = |V_{ub}| f_+(0)=  (7.2\pm 1.8) \times 10^{-4} \,.
 \qquad
\end{equation}
This estimate of 25\% uncertainty accounts for the 10\% experimental
uncertainty, and $\sim 20\%$ theory uncertainty from perturbative and power
corrections.  The experimental uncertainty includes $\gamma=70^\circ \pm
15^\circ$ which covers the range from global fits and that preferred by the SCET
based $B\to\pi\pi$ method from Ref.~\cite{gamma}.  As noted in~\cite{Bpipi} the
dependence of $|V_{ub}|f_+(0)$ on $\gamma$ is mild for larger $\gamma$'s.
Estimates for perturbative and power corrections to Eq.~(\ref{Vf}) are each at
the $\sim 10\%$ level even when ``chirally enhanced'' terms are
included~\cite{chisens,BBNS}.

Next we consider lattice QCD input points, $f^k_{\rm in}$, which are crucial in
fixing the form factor normalization.
%\footnote{ 
Technically,
  using staggered fermions might add model dependence from the $(\det M)^{1/4}$
  trick. We take the remarkable agreement in~\cite{pLQCD} as an indication that
  this model dependence is small.%} 
  Using the unquenched MILC configurations, Refs~\cite{FNAL,HPQCD} 
  find consistent results with different heavy quark actions.
  As our default we use the Fermilab results since they have a point at larger
  $q^2$:
\begin{eqnarray} \label{flatt}
&&%\hspace{-0.1cm}
 \quad  f^1_{\rm in} = f_+(15.87) = 0.799 \pm 0.058 \pm 0.088 \,,
   \qquad\quad \\ 
&&\quad f^2_{\rm in} = f_+(18.58) = 1.128 \pm 0.086 \pm 0.124 \,,
   \qquad\quad \nn\\ 
&&\quad  f^3_{\rm in} = f_+(24.09) = 3.262 \pm 0.324 \pm 0.359 \,.
   \qquad\quad \nn
\end{eqnarray}
The first errors in (\ref{flatt}) are statistical, $\pm \sigma_i$, and the
second are $11\%$ systematic errors, $\pm yf^i_{\rm in}$, with $y=0.11$.  For
the lattice error matrix, we use $E_{ij}= \sigma_i^2 \delta_{ij}+ y^2 f^i_{\rm
  in} f^j_{\rm in}$, which takes $\sigma_i$ uncorrelated and includes 100\%
correlation in the systematic error. Of the eleven reported lattice points we
use only three at separated $q^2$. This maximizes the shape information while
minimizing additional correlations that may occur in neighboring points, for
example from the chiral extrapolation.

Chiral perturbation theory (ChPT) gives model independent input for $f_+$ (and
$f_0$) when $E_\pi\sim m_\pi$, namely
\begin{equation}
  f_+\big(q^2(E_\pi)\big)\!=\!
  \frac{gf_Bm_B}{2f_\pi(E_\pi\!+\!m_{B^*}\!-\!m_B)} 
  \Big[ 1 \!+ {\cal O}\Big(\frac{E_\pi}{\Delta}\Big) \Big],
\end{equation}
where $g$ is the $B^*B\pi$ coupling and $f_B$ the decay constant.  Possible pole
contributions from the low lying $J^\pi=0^+,1^+,2^+$ states vanish by parity and
angular momentum conservation. The first corrections scale as $E_\pi/ \Delta$,
where $\Delta\sim 600\,{\rm MeV}$ is the mass splitting to the first radially
excited $1^-$ state above the $B^*$.  We take $g=0.5$.  This is compatible with
$D^*$ decays using heavy quark symmetry. Updating the ChPT fit
in~\cite{Iain} by including both $\Gamma(D^{*+})$ and $D^*$ Br-ratios, gives
$g_{D*D\pi}\simeq 0.51$ (at an order where there are no counterterm operators
and no $1/m_c$ corrections absorbed in $g$).  For the lattice average
Hashimoto~\cite{Hashimoto} gives $f_B=189\,{\rm MeV}$.  Thus,
\begin{eqnarray} \label{f4}
  f^4_{\rm in} = f_+(26.42) = 10.38 \pm 3.63 \,,
\end{eqnarray}
where this fairly conservative $35\%$ error is from uncertainty in $g f_B$, and
from the $m_\pi/\Delta\sim 23\%$ corrections.

\begin{figure}[t!]
  \centerline{ 
  \epsfxsize=8.2truecm \hbox{\epsfbox{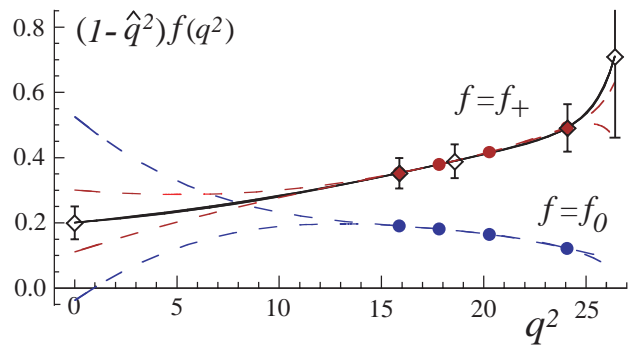}} 
   }
  \vskip -0.4cm 
  {\caption[1]{ Upper and lower bounds on the form factors from dispersion
      relations, where $\hat q^2=q^2/m_{B^*}^2$ and the $(1\!-\!\hat q^2)$
      factor removes the $B^*$ pole. The overlapping solid black lines are
      bounds $F_{\pm}$ derived with the SCET point, 3 lattice points, and the
      ChPT point (diamonds with error bars). The dashed lines are the bounds
      derived using instead four lattice points (shown by the dots). Input point
      errors are not included in these lines, and are analyzed in the text.
    }\label{fig1}}
\vskip -0.1cm
\end{figure}

{\em Determining $f_+$}. To determine $f_+(t)$ we drop $a_{k\ge 6}$ in
Eq.(\ref{ff}), and take $a_5\to a_5(1-z^2)^{-1/2}$ which properly bounds the
truncation error~\cite{BL}.  The $f^{0-4}$ input points then fix $a_{0-4}$ as
functions of $a_{5}$.  Functions that bound $f_+(t)$ are determined
from the maximum and minimum values of $a_{5}$ satisfying~(\ref{sum}) with
$n_A=5$. Thus we solve
\begin{eqnarray} \label{eqsolve}
% t = 0.65 tminus:
&&\hspace{-0.4cm}
18.3 a_0 \!+\! 3.96 a_1 \!+\! 0.857 a_2 \!+\! 0.185 a_3 \!+\! 0.0401 a_4 
     \\
 && %\hspace{-0.3cm}
   \!+\! 0.00887 a_5 = {f^0}/{|V_{ub}|}\,, \nn\\
&&\hspace{-0.4cm}
  37.8 a_0 \!+\! 0.960 a_1 \!+\! 0.0244 a_2 \!+\!  0.000619 a_3 \!+\!
   1.57\!\times\!\! 10^{-5} a_4  \nn\\
 &&
  \!+ 4.00\!\times\!\!10^{-7} a_5  = f^1 \,, \ \ldots \,, \nn\\
%&&\hspace{-0.4cm}
%  48.2 a_0 \!-\! 1.48a_1 \!+\! .0452 a_2 \!-\! 0.00138 a_3 \!+\!
%   4.34\times 10^{-5} a_4 \!-\! 1.30\times 10^{-6} a_5
%    \! =\! f_2 \,,\nn\\
%&&\hspace{-0.4cm}
%  119.4 a_0 \!-\! 24.7 a_1 \!+\! 5.09 a_2 \!-\! 1.05 a_3 \!+\! 0.217 a_4 
%    \!-\! .0457 a_5 \! =\! f_3 \,,\nn\\
&&\hspace{-0.4cm}
  304.0 a_0 \!-\! 103.6 a_1 \!+\! 35.3 a_2 \!-\!12.0 a_3 \!+\! 4.10 a_4 
    \!-\! 1.49 a_5 \! =\! f^4 ,\nn\\
&& \hspace{-0.4cm}
  a_0^2 + a_1^2 + a_2^2 + a_3^2 + a_4^2 + a_5^2 = 1 \,. \nn 
\end{eqnarray}
In Eq.(\ref{ff}) this yields two solutions, $F_{\pm}$, with parameters
\begin{eqnarray} \label{fp}
 f_+(t) = F_{\pm}(t,\{f_0/|V_{ub}|, f_{1}, f_2, f_3, f_4\}) \,.
\end{eqnarray}
To see how well these solutions bound the form factor we fix $|V_{ub}|=3.6\times
10^{-3}$, $f^i=f^i_{\rm in}$ and plot the bounds as the two black solid lines in
Fig.~\ref{fig1}.  The curves lie on top of each other. For comparison we show
dashed lines for the bounds on $f_+$ and $f_0$ obtained using four lattice
points (shown as dots).  With these inputs the constraint $f_+(0)=f_0(0)$ is
less effective than using the SCET point.

{\em $|V_{ub}|$ from total {\rm Br}-fraction}. Equating Eq.(\ref{avg}) with the
theoretical rate obtained using Eqs.(\ref{fp})  gives an
analytic equation for $|V_{ub}|$. With $f^i=f_{\rm in}^i$ the solution is
\begin{eqnarray} \label{Vub1}
 |V_{ub}|= (4.13 \pm 0.21\pm 0.58) \times 10^{-3} \,.
\end{eqnarray}
The first error is experimental, 5.2\%, propagated from Eq.(\ref{avg}). The
second error, 14\%, is from theory and is broken down in Table~\ref{table}. It
is dominated by the input points.  The bound uncertainty from the choice of
solution is $<1\%$ (but would grow to $\pm 12\%$ without the SCET point).  The
error from $m_b$ and the order in the OPE and are very small because shifts in
the normalization through $\chi_{f_+}^{(0)}$ are compensated by shifts in the
$a_n$ coefficients, except for the last term $a_5$ which gives a small
contribution.  To ensure consistency with the dispersion bounds the input point
uncertainty is calculated using the Lellouch-method of generating random points
from Gaussians~\cite{Lellouch}, giving $10^3|V_{ub}|=(3.96\pm 0.20 \pm 0.56)$.
Our distributions were determined using Eqs.~(\ref{f0},\ref{flatt},\ref{f4}) and
the correlation matrix $E_{ij}$.  Taken individually the SCET and ChPT points
give $\sim 5\%$ error, so the lattice uncertainty dominates.
\begin{table}[t!] 
%add [H] placement to break table across pages
\begin{ruledtabular}
\begin{tabular}{cccc}
 Type of Error & Variation From &
   $\delta |V_{ub}|^{\rm Br} $ & $\delta |V_{ub}|^{q^2} $ \\
\hline
 Input Points & 1-$\sigma$ correlated errors & $\pm 14\%$ & $\pm 12\%$ \\
 Bounds & $F_+$ versus $F_-$ & $\pm 0.6\%$ & $\pm 0.04\%$ \\
 $m_b^{\rm pole}$ & $ 4.88 \pm 0.40$ &  $\pm 0.1\%$ &  $\pm 0.2\%$\\
 OPE order &  2 loop\:$\to$\:1 loop  & $-0.2\%$ & $+0.3\%$\\
%\hline 
\end{tabular}
\end{ruledtabular}
\vskip -0.2cm 
\caption{\label{table}
Summary of theoretical uncertainties on $|V_{ub}|$.  Results are shown for  
an analysis from the total branching fraction, $\delta|V_{ub}|^{\rm Br}$, and 
from using the $d\Gamma/dq^2$ spectrum, $\delta |V_{ub}|^{q^2}$. For the input
point error we quote the average from $F_\pm$.
}
\vskip -0.2cm
\end{table}

{\em $|V_{ub}|$ from $q^2$ spectra}. Results for partial branching fractions,
$({\rm Br}_i^{\rm exp}\pm \delta {\rm Br}_i)$, over different bins in $q^2$ are
also available.  Cleo~\cite{Cleo} and Belle~\cite{Belle} present results for 3
bins with untagged and $\pi^+$ semileptonic tags respectively.
Babar~\cite{Dingfelder} recently presented total rates from hadronic \& leptonic
$\pi^+$ and $\pi^0$ tags as well as $\pi^+$ semileptonic tagged data in 3-bins
and untagged data over 5-bins.  By fitting to these 17 pieces of data with
Minuit we exploit the $q^2$ shape information. To do this we define
\begin{eqnarray} \label{chi2}
 \chi^2\!\! &=&\!\!
  \sum_{i=1}^{17} \frac{ [{\rm Br}_i^{\!\!\rm exp}\!
    - {\rm Br}_i(V_{ub}, F_{\pm}) ]^2} {(\delta {\rm Br}_i)^2}
   + \frac{[f^0_{\rm in}-f^0]^2}{(\delta f^0)^2} \
  \\
&+&  \frac{[f^4_{\rm in}-f^4]^2}{(\delta f^4)^2}
 + \sum_{i,j=1}^{3}{\big[f^i_{\rm in}-f^i\big]\big[f^j_{\rm in} - f^j\big]}
   (E^{-1})_{ij} 
   \,,\nn
\end{eqnarray}
and minimize $\chi^2$ as a function of $|V_{ub}|$ and $f^{0-4}$. $\chi^2$
contains both experimental and theoretical errors, with $E^{-1}$ the inverse
error matrix. By allowing $f^{0-4}$ in $F_{\pm}$ to move away from $f_{\rm
  in}^{0-4}$ the theoretical rate is allowed to adjust itself based on the
experimental $q^2$ shape. 

\begin{figure}[t!]
  \centerline{ 
  \epsfxsize=7.8truecm \hbox{\epsfbox{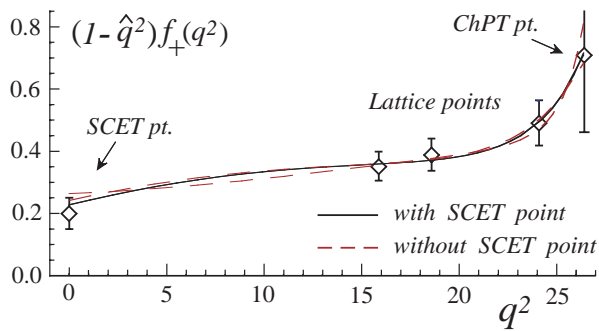}} 
   }
  \vskip -0.4cm 
  {\caption[1]{Results from the $\chi^2$ fit of $|V_{ub}|$ and $f^{0-4}$ to the
      $q^2$ spectra ($\hat q^2=q^2/m_{B^*}^2$).  The two solid lines are
      obtained using either the $F_+$ or $F_-$ solutions from Eq.~(\ref{fp}).
      The two dashed lines repeat this analysis without using the SCET point.
    }\label{fig2a} }
\vskip -0.2cm
\end{figure}
\begin{figure}[t!]
  \centerline{ 
  \epsfxsize=7.5truecm \hbox{\epsfbox{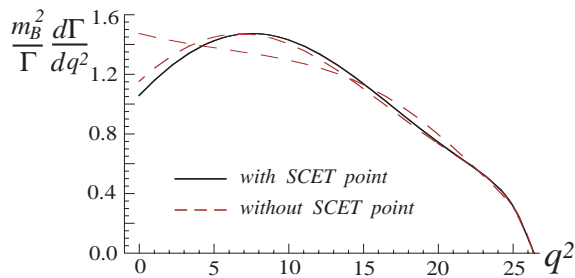}} 
   }
  \vskip -0.2cm 
  {\caption[1]{The curves are as in Fig.2, but for the decay rate.}\label{fig3a}}
\vskip -0.3cm
\end{figure}
Minimizing (\ref{chi2}) gives $\chi^2/(dof) = 1.04$ and
\begin{eqnarray} \label{Vub2}
 |V_{ub}|= (3.54 \pm 0.47 ) \times 10^{-3} \,.
\end{eqnarray}
Results for $f_+(q^2)$ and $d\Gamma/dq^2$ are shown by the black solid curves in
Figs.~\ref{fig2a} and~\ref{fig3a}.  Eq.(\ref{Vub2}) has a total error of 13\%.
If we fix $f^{0-4}=f_{\rm in}^{0-4}$ then the experimental error is $4.9\%$, ie.
$\delta |V_{ub}|=\pm 0.17$. The remainder, $\delta |V_{ub}|=\pm 0.44$ is from
the input points, so the $q^2$ spectra brought this theory error down to $12\%$.
Other uncertainties are small as shown in Table~\ref{table}. The experimental
spectra favor a larger form factor between the lattice and SCET points. This
decreases the value of $|V_{ub}|$ from that in (\ref{Vub1}).  Using
Eqs.~(\ref{f0},\ref{f4}) this fit yields
\begin{eqnarray}
  f_+(0) = 0.227 \pm 0.047 \,,\quad
  g\, f_B = 96 \pm 29 \, {\rm MeV} \,,
\end{eqnarray}
consistent with our inputs. This $f_+(0)$ has 21\% error.

If we entirely remove the SCET point $f^0$ from Eq.(\ref{chi2}) then we obtain a
fit that uses only semileptonic data, shown by the dashed red lines in
Figs.~\ref{fig2a} and~\ref{fig3a}. The spectrum is now determined less precisely
at small $q^2$, since this data only bounds the area in the smallest $q^2$-bin.
The result is $|V_{ub}|=(3.56\pm 0.48)\times 10^{-3}$. It has the same input
point error as Eq.(\ref{Vub2}) and a somewhat larger bound error,
$\delta|V_{ub}|=1.8\%$.  Turning the use of Eq.(\ref{f0}) around, we can combine
it with $f_+(0)$ to get an independent method of fixing $|V_{ub}|$ from the
nonleptonic data. The semileptonic fit gives $f_+(0)=0.25\pm 0.06$, so
Eq.(\ref{f0}) yields $|V_{ub}|^{nonlep} = (2.9 \pm 1.0) \times 10^{-3}$.

Our final result for $|V_{ub}|$ is given in (\ref{Vub2}). The final theory error
is dominated by the lattice points, and is very close to their error. It will
decrease with this error in the future. See also~\cite{Flynn}. To go beyond the
analysis here it will be interesting to study the additional error correlation
implied by the dispersion relations when lattice input points are included that
are closer together.

We thank J.Branson, J.Flynn, L.Gibbons, A.Kronfeld, M.Okamoto, D.Pirjol,
and J.Shigemitsu for helpful conversations.  This work was supported by the
U.S.\ Department of Energy under DOE-FG03-97ER40546 (B.G.), DOE-ER-40682-143 and
DEAC02-6CH03000 (I.R.), the cooperative research agreement DF-FC02-94ER40818 and
Office of Nuclear Science (C.A. and I.S.), and a DOE OJI award and Sloan
Fellowship (I.S.). I.S. thanks the Institute of Nuclear Theory for their
hospitality during the completion of this work.

\end{document}